\documentclass[sigplan,10pt,nonacm]{acmart}
\renewcommand\footnotetextcopyrightpermission[1]{}
\usepackage{booktabs}

\usepackage{subcaption}

\usepackage{graphicx}
\usepackage{listings}
\usepackage{booktabs}
\usepackage[linesnumbered,ruled,noend,figure]{algorithm2e}
\usepackage{caption}
\usepackage{subcaption}
\usepackage{amsmath}
\usepackage{fancyhdr}
\usepackage{verbatim}
\usepackage{CJKutf8}
\usepackage{enumitem}
\usepackage{makecell}

\newcommand{\oursys}{SMEPilot}

\begin{document}
\begin{CJK*}{UTF8}{gbsn}

\hyphenpenalty=600

\lstset{
  language=Python,
  basicstyle=\fontsize{8}{8.5}\ttfamily,
  keywordstyle=\color{blue},
  commentstyle=\itshape\color{green!40!black},
  frame=single,
  numbers=left,
  stepnumber=1,
  emph={getGridDim, getBlockDim, kernel, getBlockIdx, StreamModule, virtual, Schedule, Signal, GetBlockExecutorId},
  emphstyle=\textbf,
  morekeywords={class, void, size_t, vector, Map, sEUType},
  deletekeywords={bool},
  emph={[2] StreamModule, virtual, Schedule, Signal, GetBlockExecutorId},
  emphstyle={[2]\color{purple!80!black}}
}

\date{}

\title{\oursys{}: Characterizing and Optimizing LLM Inference with Scalable Matrix Extensions}

\fancyhead{}
\renewcommand\footnotetextcopyrightpermission[1]{}

\author{Feiyang Chen, Haibo Chen}
\affiliation{
  \institution{IPADS, Shanghai Jiao Tong University}
  \country{}
}

\begin{abstract}

Modern CPUs increasingly integrate matrix extensions, such as Arm
Scalable Matrix Extension (SME), that provide high-throughput matrix
execution within the CPU.  For LLM inference, however,
these units are not a universal replacement for conventional CPU cores:
prefill, decode, attention, and KV-cache operations expose different
arithmetic intensities, vector behavior, and layout requirements, while
SME units and CPU cores still compete for shared memory bandwidth.  This
paper studies this mismatch through a roofline-based characterization of
SME-enabled CPUs and uses the resulting model to guide operator-level
execution choices.  We present \oursys{}, an LLM inference engine that
selects CPU-only, SME-only, or cooperative SME+CPU execution for each
operator shape.  \oursys{} partitions matrix work across SME and CPU
cores at tile granularity, overlaps SME-suitable matrix stages with
CPU-suitable vector stages in attention, and maintains layout state so
packed tensor representations are reused rather than repeatedly rebuilt
on critical paths.  Across Llama-3.2-3B, Qwen3-4B, and Qwen3-30BA3B on
phone, PC, and server platforms, \oursys{} improves end-to-end inference
performance by up to 3.94$\times$.

\end{abstract}

\settopmatter{printfolios=true}
\maketitle

\sloppy

\section{Introduction}
\label{sec:intro}

The rapid progress of large language models (LLMs) has made generative
AI a common component of interactive applications, from chat assistants
and AI agents to multimodal generation~\cite{minaee2024llmsurvey,luo2025llmlagent,zhao2026unifiedmultimodal}.
As these applications span platforms from edge devices to data centers,
CPUs continue to play a critical role in executing diverse AI tasks
owing to their widespread availability and cost efficiency~\cite{gerganov2023llamacpp,wei2025tmac,xu2026arclight}.

At the same time, the CPU itself is evolving for LLMs. Modern CPUs are
adding matrix extensions, such as Arm Scalable Matrix Extension (SME)~\cite{arm2024smeintro},
that provide high-throughput
matrix instructions inside a conventional CPU package.
They add
specialized matrix registers and instructions for computing 2D tiled matrix
operations, such as outer products, directly
on the CPU.
These extensions promise a middle ground between GPU-style accelerators and ordinary
vector cores: they expose more matrix throughput without requiring the
application to leave the CPU, reusing its easy and mature ecosystem.

Prior work has studied how to use Scalable Matrix Extensions to accelerate
specific kernels. For example, ARM SME has been characterized for dense
matrix multiplication~\cite{deng2025demystifying} and
sparse matrix multiplication~\cite{lei2025loops}.

However, they do not study how to use SME systematically across the entire
LLM inference process. In particular, it remains unclear how a runtime
should schedule matrix extensions and CPU vector cores across inference
phases such as prefill and decode, and across operators such as FFN, MoE and
attention.

This paper presents the first comprehensive characterization on CPUs with scalable matrix extensions for LLM
inference.  To reason about when SME units should be used, we build a unified
roofline-based model that captures both the distinct computing throughput of
SME and CPU cores and their shared memory bandwidth.  For each
LLM operator and shape, the model estimates whether execution is
compute-bound, memory-bound, or in a backend-asymmetric ridge-crossing
region, and uses this classification to predict whether the operator
should run on SME units, CPU cores, or both.

We identify that treating SME as a drop-in replacement for CPU matrix kernels
is insufficient.
There are two primary reasons for this. First, for large matrix-heavy operators
in prefill phases, SME and CPU cores can both contribute to throughput,
and using only one backend leaves useful hardware idle.
Second, different models and input requests can shift an operator between
SME-friendly and CPU-friendly regimes, which causes a single fixed kernel replacement
to underperform across the inference process.

To address this problem, we introduce \oursys{}, an accelerated inference engine for CPUs with scalable
matrix extensions.  \oursys{} provides a unified framework that schedules SME and CPU cores
across prefill and decode phases,
and across operators with different shapes and hardware affinities.
For each inference request, \oursys{} uses
the roofline classification to generate an execution plan that schedules
CPU-only, SME-only, or mixed SME+CPU execution for each operator.

After the execution plan is generated, \oursys{} provides a performant runtime
to execute the plan on SME units and CPU cores.  However, attaining high
performance requires more than choosing the right plan:
the runtime must preserve parallelism, hardware utilization, and
data locality while executing the plan.  We identify three execution gaps
that prevent the roofline-guided plan from realizing its predicted
benefit.
\begin{itemize}
    \item \textbf{Spatial utilization gap.}
    A policy that assigns each operator entirely to a single execution target
can leave useful hardware idle.  Large
    GEMM operators expose enough arithmetic intensity to benefit from
    both SME and CPU cores, but assigning the entire operator to only one
    side wastes the throughput of the other.

    \item \textbf{Temporal bubble gap.}
    LLM inference operators, especially attention, interleave phases with
    different hardware affinity.  Matrix multiplication such as $QK^\top$
    can benefit from SME, while softmax, masking, reductions, and other
    vector-heavy phases are better suited to CPU cores.  If these phases are
    executed sequentially at coarse operator boundaries, one hardware unit
    waits while the other runs, creating bubbles that are invisible to a
    per-operator roofline model.

    \item \textbf{Layout compatibility gap.}
    SME kernels require packed, tile-friendly matrix layouts to expose their
    throughput, whereas the rest of the LLM pipeline often consumes and
    produces conventional tensor layouts.  Naively packing inputs before
    every SME invocation adds memory traffic and disrupts data locality,
    reducing the effective arithmetic intensity assumed by the roofline
    model.

\end{itemize}

To bridge these gaps between the roofline-guided plan and runtime execution,
\oursys{} adopts three techniques.
\begin{itemize}
    \item \textbf{Tile-level work partitioning.}
    To bridge the spatial utilization gap, \oursys{} partitions matrix work
    across SME and CPU cores at tile granularity.  This allows both
    backends to contribute within the same operator while respecting the
    tile shapes required by their optimized kernels.

    \item \textbf{Phase-aware pipeline execution.}
    To bridge the temporal bubble gap, \oursys{} pipelines SME-friendly
    matrix phases with CPU-friendly vector phases.  This lets attention
    phases such as matrix multiplication and softmax/reduction overlap
    when their dependencies allow, reducing idle time on both sides.

    \item \textbf{Layout-aware runtime.}
    To bridge the layout compatibility gap, \oursys{} tracks tensor layouts
    as runtime state and performs layout-aware graph conversion.  Static
    weights are packed once off the request critical path, while online
    activations such as KV-cache entries are packed at the producer side
    when downstream SME consumers can reuse them.  This avoids repeated
    packing and wasted memory bandwidth on SME critical paths.
\end{itemize}

\oursys{} delivers state-of-the-art performance on Phone, PC, and
server CPU platforms. In end-to-end evaluations,

\oursys{} achieves up to 3.94$\times$ end-to-end speedup
over llama.cpp~\cite{gerganov2023llamacpp} across
the evaluated device-model-workload configurations, including both dense models and MoE models.
For prefill phases, \oursys{} improves FFN GEMM by 1.42--1.67$\times$ over
highly optimized Apple Accelerate~\cite{apple2026accelerate}(SME)
and 6.06--7.43$\times$ over GGML~\cite{gerganov2023ggml}(CPU),
while
improves attention by 1.56--3.50$\times$ over a FlashAttention CPU~\cite{dao2022flashattention}.
These gains come from \oursys{}'s ability to use SME and CPU cores together for large matrix-heavy prefill operators.
For decode phases, \oursys{} achieves up to 3.48$\times$ speedup over CPU-only
inference. This demonstrates that \oursys{} can still improve
hardware utilization in regimes with less computation and stronger memory
bandwidth pressure.

\section{Preliminaries}
\label{sec:moti}

\subsection{LLM Inference}

Modern LLMs are commonly built from Transformer blocks,
where each layer alternates between attention and feed-forward network
(FFN) computation~\cite{Ashish17AttentionIsAllYouNeed}.
From a system perspective, these blocks contain a mixture of operator types.
Both of the attention and FFN blocks contains two kinds of operators:
matrix operations such as matrix multiplication,
and non-linear operations such as softmax, normalization, and activation functions.
This mixture makes CPU matrix extensions
promising for LLM inference, because they can accelerate the matrix-heavy
parts, but also non-trivial to use, because not every phase benefits from
matrix throughput.

LLM inference also has two execution phases with different shapes:
prefill and decode.  During prefill, the model processes all prompt tokens
at once and produces KV-cache entries for them~\cite{kwon2023vllm,agrawal2024sarathiserve,zhong2024distserve}.
The token dimension is
therefore large, which increases arithmetic intensity and exposes
substantial matrix parallelism in attention and FFN operators.  During
decode, the model usually processes one new token per step, or a small
number of tokens when using batching~\cite{yu2022orca} or multi-token prediction~\cite{gloeckle2024multitoken,cai2024medusa}.
These operators are therefore more memory-bound:
they read large weight and KV-cache matrices but perform little computation per request~\cite{mutlu2018bottlenecks}.
Therefore, during LLM inference, different computation characteristics
occurs due to the different phases and different operators, which
poses challenges for utilizing matrix extensions efficiently.

\subsection{Scalable Matrix Extension}

The rising demands of AI workloads on both cloud and edge platforms
have led to matrix extensions being added to modern CPU architectures.
Arm Scalable Matrix Extension (SME) is an Armv9-A architecture
extension that adds matrix-oriented execution support to the
CPU~\cite{arm2024smeintro,arm2026acle}.
SME introduces a specialized outer-product instruction and
a new matrix register array storage,
significantly increasing the matrix throughput of Arm CPUs~\cite{remke2024hellosme,deng2025demystifying}.

SME has been supported in a range of CPUs.
Apple's M4 was
the first publicly available chip reported to support SME, and M4-family
CPUs are now used across Apple mobile and laptop platforms~\cite{remke2024hellosme}.
Recent flagship mobile Arm platforms are also adopting SME.
For example,
MediaTek's Dimensity 9500 introduces Armv9.3 support with SME2 instructions,
and Qualcomm Snapdragon 8 Elite Gen 5 uses third-generation Oryon
CPU cores that also support SME.
On the cloud side,  Arm-based server CPUs such as newer Kunpeng 920 also support SME.

\section{Roofline Characterization of SME-enabled CPUs}
\label{sec:sme}

Scalable Matrix Extensions reshape the performance structure of CPU inference.
In order to systematically analyze the performance of different LLM operators
on the CPU with SME,
we design a unified roofline model that can capture the performance characteristics
of both SME and CPU cores.
Based on this roofline model, we characterize the performance of different LLM operators on the CPU with SME,
and derive the design principles for \oursys{}.

\begin{figure}[t]
\centering
\includegraphics[width=0.98\columnwidth]{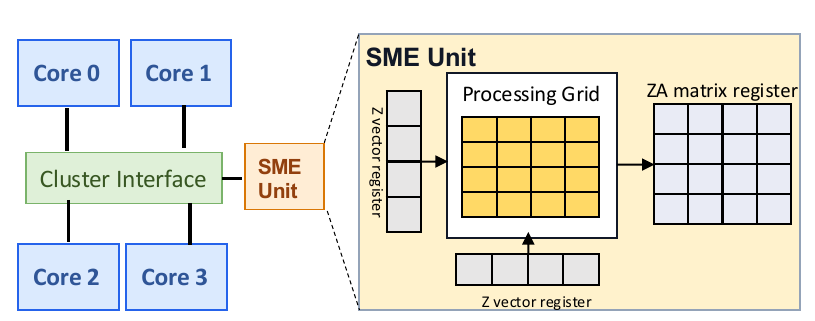}
\caption{The hardware design of CPU with SME. Left: an CPU
cluster attaches ordinary CPU cores and an SME unit.
Right: the SME unit executes outer products instruction from
Z vector registers and output to ZA matrix registers.}
\Description{A two-panel diagram introducing Arm SME. The left panel
shows CPU cores connected to a shared DSU and memory hierarchy, with an
SME unit attached to the same cluster. The right panel shows an SME
engine datapath with decoder, SVE execution, predicate registers, vector
registers, a processing grid, ZA tile array storage, ZA extraction, and
load/store execution.}
\label{fig:sme-arch}
\end{figure}

\begin{figure*}[t]
\centering
\includegraphics[width=1\textwidth]{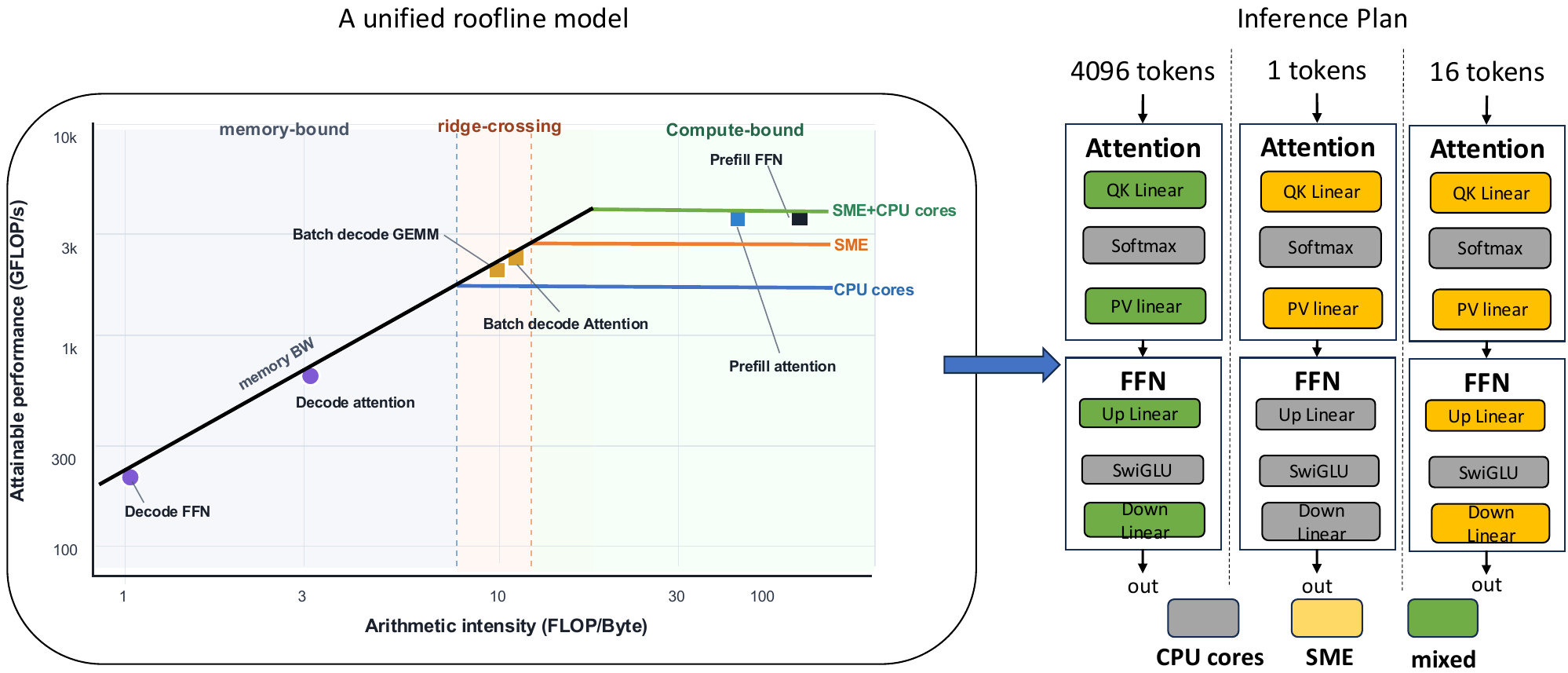}
\caption{Left: SME+CPU roofline view of representative LLM operators.
CPU cores, SME, and mixed SME+CPU execution share the same memory bandwidth
but expose different compute peaks, creating different ridge points:
memory-bound, compute-bound and ridge-crossing.
Right: The inference plan for different input shapes.
Operators are placed into different backends
based on their position estimated by the unified roofline model.}

\label{fig:roofline-sme-cpu}
\end{figure*}

\subsection{SME and CPU Cores: Additive Compute, Shared Bandwidth}

Although SME instructions are exposed through the CPU programming
model, the underlying execution resources are distinct from ordinary CPU vector
cores, as shown in Figure~\ref{fig:sme-arch}.
SME provides specialized matrix state and outer-product instructions for
high-throughput tiled matrix computation, while CPU cores continue to provide
general-purpose scalar and vector execution.

Two properties determine how these resources should be used for LLM inference.

\begin{figure}[t]
    \centering
    \includegraphics[width=\columnwidth]{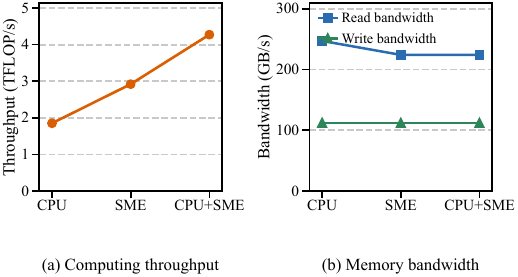}
    \caption{The computing and memory characteristics of SME and CPU cores.
    SME and CPU cores provide near-additive matrix throughput but
    share memory bandwidth.
    }
    \Description{Two side-by-side line charts over SME, CPU cores, and
    mixed SME plus CPU execution. The left chart shows computing throughput
    increasing for mixed execution, while the right chart shows read and
    write memory bandwidth remaining similar across the three configurations.}
    \label{fig:char-additive}
    \label{fig:char-shared}
\end{figure}

\textbf{Near-additive matrix throughput.}
For matrix-heavy operators with enough independent work, SME units and CPU
cores can both contribute useful computation. As shown in Figure~\ref{fig:char-additive}, two SME
workers alone reach 2.92 TFLOP/s on FP16 matrix work and eight CPU-core workers
reach 1.85 TFLOP/s. Running them together reaches 4.27 TFLOP/s, which is close
to the sum of the individual throughputs. This indicates that SME and CPU cores
should not be treated as mutually exclusive backends for large matrix operators.

\textbf{Shared memory bandwidth.}
The same additivity does not hold for memory traffic. SME units and CPU cores
share the cache hierarchy and memory controller. As shown in Figure~\ref{fig:char-shared}, CPU cores
reach 247 GB/s read bandwidth, SME reaches 224 GB/s, and mixed SME+CPU execution
does not increase the memory roof. All three configurations sustain similar
write bandwidth. Thus, adding SME raises the compute ceiling but does not raise
the memory-bandwidth roof.

These two properties make SME-enabled CPUs different from both homogeneous
multicore CPUs and discrete accelerators. Compute resources can be added within
one operator, but memory bandwidth remains coupled across them.

\subsection{Multi-Ceiling Roofline Model}

The coupled structure of SME and CPU cores
creates a placement problem that a conventional
roofline model does not capture.  A conventional roofline view can
classify whether an operator is compute-bound or memory-bound on a single
hardware target, but LLM inference on SME-enabled CPUs must choose among
CPU-only, SME-only, and mixed SME+CPU execution. We therefore build a
coupled roofline model with a shared memory-bandwidth roof and multiple
compute ceilings for CPU cores, SME units, and mixed SME+CPU execution.

For an operator $o$, let $W_o$ denote its floating-point work and $Q_o$ denote
its useful memory traffic. Its arithmetic intensity is
$I_o = W_o / Q_o$. For each execution target
$h \in \{\mathrm{CPU}, \mathrm{SME}, \mathrm{MIX}\}$, let $P_h$ denote the
measured compute throughput. The mixed target represents concurrent SME+CPU
execution and has a compute ceiling close to
$P_{\mathrm{SME}} + P_{\mathrm{CPU}}$. Since all targets share the same memory
system, they use a common memory bandwidth $B$.

The attainable performance of target $h$ is modeled as
\[
R_h(o) = \min(P_h, B I_o).
\]

The ridge point of target $h$ is
\[
I_h^{ridge} = \frac{P_h}{B}.
\]
Because $P_{\mathrm{CPU}} < P_{\mathrm{SME}} < P_{\mathrm{MIX}}$ while $B$ is
shared, the model contains one memory roof but multiple compute ceilings.

As shown in Figure~\ref{fig:roofline-sme-cpu},
unlike a traditional roofline model with only one ridge point,
this multi-ceiling roofline model partitions the performance space into three regimes:
memory-bound, compute-bound, and backend-asymmetric ridge-crossing.
These three regimes correspond to three different placement plans for LLM operators on SME-enabled CPUs.

\textbf{Memory-bound regime.}
Operators below the CPU ridge point are memory-bound on all targets. Since SME
does not increase memory bandwidth, moving these operators to SME cannot reduce
the dominant cost.
This regime includes single-token decode GEMV and
decode attention, where the operator reads large weight or KV-cache matrices
but performs little computation per request.

CPU execution is preferable for this region because
we observe that SME kernels typically has
lower memory utilization than CPU cores due to
higher instruction latency and lack of speculative execution.

\textbf{Backend-asymmetric ridge-crossing regime.}
Operators between the CPU and SME ridge points are compute-bound on CPU cores
but memory-bound on SME. This regime is specific to a multi-ceiling roofline model.
SME improves performance in this region not by reaching
its peak compute throughput, but by lifting the operator above the CPU compute
ceiling until it approaches the shared memory bandwidth.
Batched decode GEMM and some grouped-query
attention shapes fall into this region.

SME-only execution is preferable for this region.
These operators favor SME execution for higher computing throughput, but
mixed SME+CPU execution provides limited additional benefit because memory
bandwidth becomes the bottleneck.

\textbf{Compute-bound regime.}
Operators above the SME ridge point are compute-bound on both SME and CPU cores. In this
region, the memory bandwidth is no longer the limiting resource, so the
near-additive compute throughput of SME and CPU cores can be exploited. Large
prefill FFN and attention matrix phases fall into this region.
These operators
benefit from mixed SME+CPU execution, provided that the runtime can utilize parallelism
within the operator.

\begin{figure}[t]
    \centering
    \includegraphics[width=0.8\columnwidth]{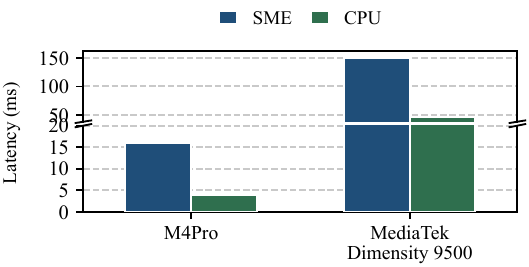}
    \caption{SME has lower throughput for non-linear vector operations.
    For a $4096 \times 4096$ fastexp kernel, CPU cores are faster than SME
    on both Apple M4 Pro and MediaTek Dimensity 9500.}
    \Description{A grouped bar chart comparing fastexp latency on Apple
    M4 Pro and MediaTek Dimensity 9500. In both device groups, the SME bar
    is higher than the CPU cores bar, showing higher latency for SME.}
    \label{fig:char-fastexp}
\end{figure}
\noindent\textbf{Non-linear operations.}
Except for the linear operations,
LLM inference also contains non-linear operations such as softmax, normalization and
activation. We always place these non-linear operations on CPU cores,
because SME has lower throughput for these operations than CPU cores.

As shown in Figure~\ref{fig:char-fastexp}, for a $4096 \times 4096$
fastexp kernel,

SME is 4.10$\times$ slower than CPU cores on Apple
M4 Pro and 3.26$\times$ slower on MediaTek Dimensity 9500 for this
non-linear kernel.
This is because SME has low throughput and higher latency for SIMD instructions,
and non-linear operations often compose of a long sequence of SIMD instructions
with strong data dependencies.

To summarize, as shown in the right panel of Figure~\ref{fig:roofline-sme-cpu},
this unified roofline model characterizes LLM operators across different
performance regimes and guides the general placement strategy for
SME-enabled CPUs.

\subsection{Scheduling Insights from the Roofline Analysis}

The coupled roofline model not only provides placement plans for different
operators, but also reveals two general insights for the runtime of
\oursys{}.

\noindent\textbf{Insight 1: Mixed execution requires intra-operator scheduling.}
Operator-level placement is too coarse for SME-enabled CPUs.
For operators in compute bound regime,
the coupled model predicts that the best target is mixed
SME+CPU execution.
However, assigning the whole operator to either backend cannot
realize this bound: CPU-only execution misses SME throughput, while SME-only
execution leaves ordinary CPU cores idle. Reaching the mixed ceiling therefore
requires the runtime to explore parallelism within one operator and
achieve high utilization for both hardware sides.

\noindent\textbf{Insight 2: Shape-dependent placement requires dynamic scheduling.}
The same operator can occupy different regions of the coupled roofline
as the request shape changes. For example, an FFN layer may be memory-bound
during single-token decode, ridge-crossing under batched decode, and
compute-bound during long-context prefill. Attention similarly shifts with
sequence length, batch size, head configuration, and KV-cache length. Therefore,
we cannot attach a fixed backend choice to each operator. Instead, the runtime
must generate a shape-aware execution plan online, selecting CPU-only, SME-only,
or mixed SME+CPU execution according to the operator's current arithmetic
intensity and available intra-operator parallelism.

These two insights distinguish \oursys{} from a drop-in SME kernel replacement.
\oursys{} does not simply decide whether an operator should use SME; it decides
which tiles should use SME and how that decision changes with the current
request shape.

\section{\oursys{} Design}
\label{sec:design}

The characterization in Section~\ref{sec:sme} identifies the placement
regimes for different operators.  Based on this characterization,
\oursys{} designs a unified runtime that turns roofline-guided placement
into a concrete, performant execution plan.  To realize intra-operator
mixed SME+CPU execution with high hardware utilization, we design two
techniques for FFN and attention operators: tile-level work
partitioning (Section~\ref{sec:tile-par}) and phase-aware pipeline
execution (Section~\ref{sec:pipeline}).  To support dynamic scheduling
while preserving data locality, we design a layout-aware runtime
(Section~\ref{sec:layout}) and an online planner
(Section~\ref{sec:planner}).

\subsection{Tile-level Work Partitioning}
\label{sec:tile-par}

\begin{figure}[t]
\centering
\includegraphics[width=0.5\textwidth]{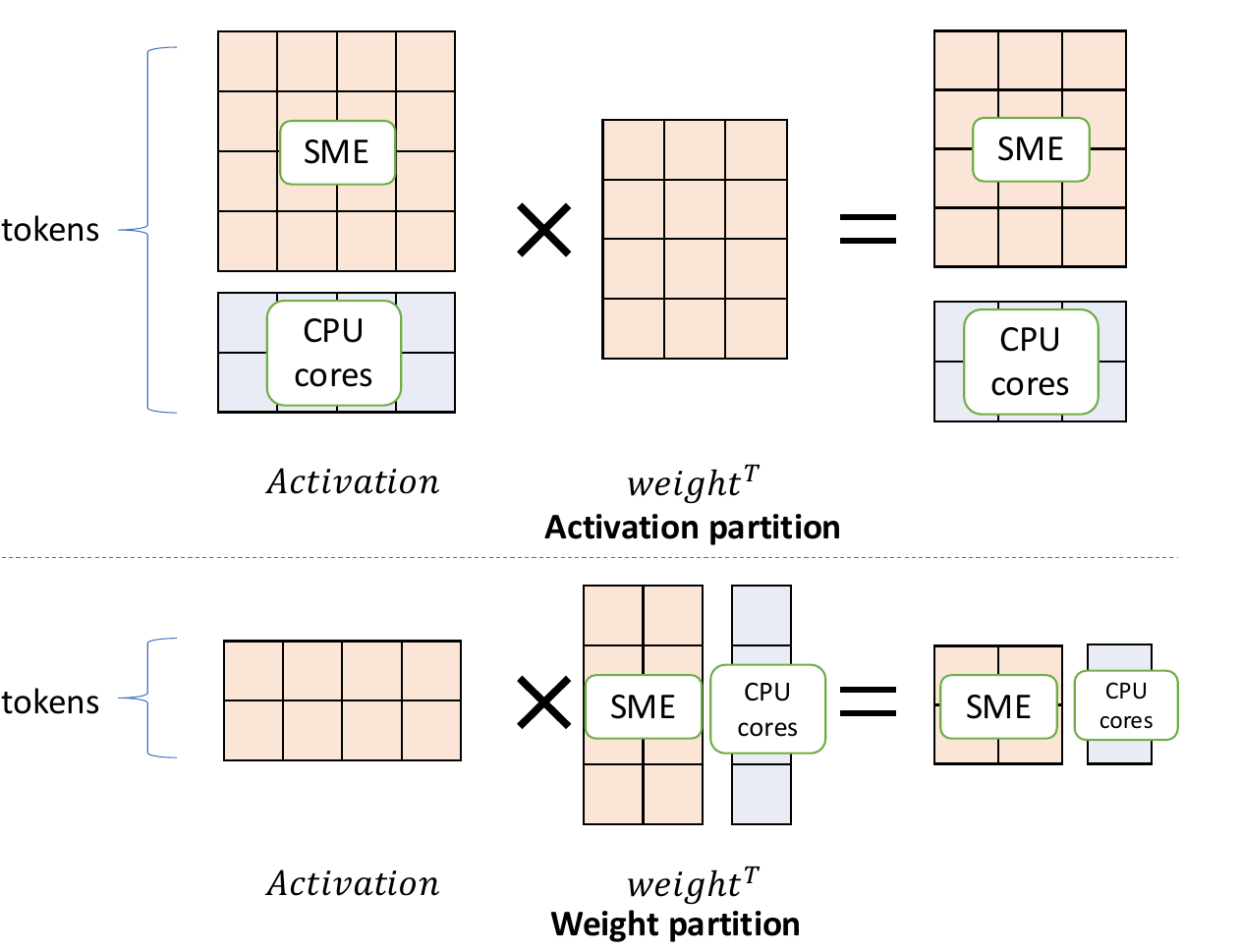}
\caption{Tile partition closes the spatial utilization gap for
compute-bound matrix operators.  The planner uses measured SME and CPU
rates, together with the worker budget and contention guard, to assign
tiles so that both backends finish at similar times.}
\Description{A matrix tile grid is split between SME tiles and CPU tiles.
A tile planner uses measured rates, worker budget, and contention guard
to dispatch work to SME and CPU worker timelines that finish at roughly
the same time.}
\label{fig:design-tile-partition}
\end{figure}

\noindent\textbf{Problem: spatial utilization gap.}
Large prefill GEMMs and FFN layers expose enough independent matrix work to
use both SME and CPU cores.
However, a whole-operator decision can choose only one
backend, while a naive split can leave the faster backend waiting for the
slower one. We need a partition plan that
achieve high utilization for both SME and CPU cores.

\noindent\textbf{Technique: partition independent output tiles.}
\oursys{} partitions a GEMM operator into output tiles and assigns a
subset of tiles to SME workers and the remaining tiles to CPU-core workers.
For a matrix multiplication $C_{M\times N}=A_{M\times K}B_{K\times N}$,
\oursys{} partitions only the output dimensions of $C$.  This choice keeps
different workers independent: each worker computes a disjoint set of output
tiles, so there is no cross-worker reduction or synchronization.  In
contrast, partitioning the reduction dimension $K$ would require partial
sums from different workers to be merged, adding synchronization and memory
traffic.

There are two natural output-dimension partition strategies, as shown in Figure~\ref{fig:design-tile-partition}.
A \emph{token-dimension partition} splits the $M$ dimension, so different
workers process different token rows.  An \emph{output-channel partition}
splits the $N$ dimension, so different workers compute different output
channels.  \oursys{} chooses the partition dimension based on the operator
shape.  When the token dimension is larger, it partitions along $M$; when
the output-channel dimension is larger, it partitions along $N$.  Splitting
the larger output dimension avoids highly elongated regions, which improves cache
locality, and makes it easier to keep each worker's region aligned with the
native tile shapes of the SME and CPU kernels.
The partition ratio is determined by the online planner(describe in Section~\ref{sec:planner}).

\noindent\textbf{MoE expert partitioning.}
Mixture-of-Experts (MoE) layers introduce another form of uneven matrix
work.  After the router assigns tokens to experts, the number of tokens
received by each expert is often skewed: some experts receive only a few
tokens, while others receive many.  \oursys{} therefore treats each expert
FFN as a separate matrix operator whose token dimension is determined by
the routed token count.  Experts with only a small number of tokens are
kept on CPU cores, because their thin matrix shapes are memory-bound and
cannot amortize SME setup, packing, and tile-alignment overheads.  Experts
with many tokens are classified by the roofline model as larger GEMM
operators and can use SME, or mixed SME+CPU execution, when their
arithmetic intensity and tile count are sufficient.  This expert-level
decision avoids forcing the same backend choice on both cold and hot
experts in the same MoE layer.

Tile-level work partitioning is used for compute-bound operators such as
prefill FFN layers and large attention matrix phases.  For memory-bound
decode GEMV-like operators, \oursys{} avoids mixed partitioning unless the
effective token width becomes large enough to expose useful parallel matrix
work.

\subsection{Phase-aware Pipeline Execution}
\label{sec:pipeline}

\begin{figure}[t]
\centering
\includegraphics[width=0.5\textwidth]{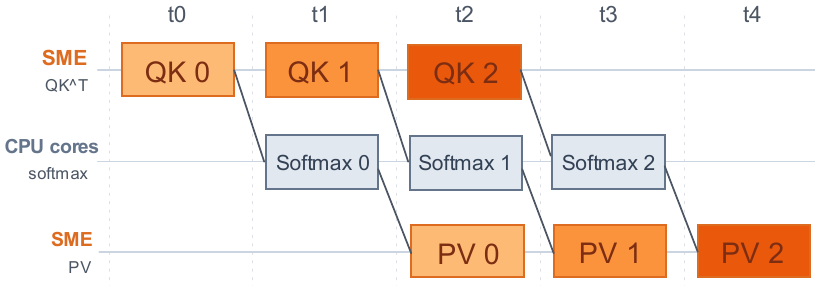}
\caption{Pipeline execution closes the temporal bubble gap in attention.
Each tile moves through SME-friendly $QK^\top$, cores-friendly
softmax, and SME-friendly $PV$ stages.  Tile-level dependencies allow
different tiles to occupy different hardware-specialized stages instead
of forcing an operator-wide barrier after each phase.}
\Description{A three-stage timeline for attention.  SME executes QK
tiles, CPU/NEON executes softmax tiles, and SME executes PV tiles.  The
same tile flows diagonally through the stages, while different tiles are
overlapped across time.}
\label{fig:attention-pipeline}
\end{figure}

\noindent\textbf{Problem: temporal bubble gap.}
Attention is not a single homogeneous matrix operator.  It contains
matrix-heavy phases, such as $QK^\top$ and $PV$, and vector-heavy phases,
such as masking, softmax, and reductions.  These phases have different
hardware affinity: the matrix phases can benefit from SME throughput,
whereas the vector-heavy phases are better executed by CPU cores.  A
coarse phase-level schedule therefore creates temporal bubbles.  If
the runtime waits for $QK^\top$ to finish before starting
softmax, CPU cores sit idle during the matrix phase; if it waits for
softmax work before starting $PV$, SME sits idle during the vector phase.
These idle periods move real execution away from the
performance predicted by the roofline model.

Existing fused attention algorithms, such as FlashAttention~\cite{dao2022flashattention},
reduce memory traffic by computing softmax online and avoiding materialized
attention matrices.  However, their primary goal is memory efficiency; they
do not by themselves address the hardware-affinity mismatch between
SME-friendly matrix phases and CPU-friendly vector phases on SME-enabled
CPUs.

\noindent\textbf{Technique: inter-phase pipelining.}
\oursys{} addresses this gap with inter-phase pipelining.  Instead of
executing attention as three operator-wide phases, \oursys{} allows
different attention tiles to occupy different phases at the same time.
Each tile still follows the dependency order $QK^\top \rightarrow$
softmax $\rightarrow PV$, but independent tiles can advance through the
pipeline concurrently.
As shown in Figure~\ref{fig:attention-pipeline},
the two matrix phases are assigned to SME workers,
while the softmax phase is assigned to CPU-core workers.  Once the
$QK^\top$ result for a tile is available, its softmax stage can start
without waiting for the entire operator to finish the $QK^\top$ phase.
This allows SME to compute the matrix phase of later tiles while CPU cores
process the vector phase of earlier tiles.

\noindent\textbf{Tile sizing.}
The tile size controls the effectiveness of the pipeline.  Very small
tiles expose more overlap opportunities, but they increase scheduling
overhead and reduce kernel efficiency.  Very large tiles improve per-kernel
efficiency, but they increase pipeline prologue and epilogue time and make the
stage times harder to balance.  \oursys{} therefore chooses tile sizes
using the offline profile and online planning: it
estimates the SME time for the matrix stages
and the CPU-core time for the softmax stage based on the
current request shape and the profiled SME and CPU throughputs, then
selects a tile shape whose
stage times are small enough to reduce long stalls while still preserving
kernel-aligned matrix tiles.

\noindent\textbf{Mask-aware matrix phases.}
Attention masks introduce another source of wasted work in the matrix
stages of the pipeline.
Causal and sliding-window masks can make entire score blocks
irrelevant, but a conventional GEMM still computes these masked elements
before discarding them.  \oursys{} uses mask-aware matrix kernels inside
the pipeline to skip tiles that are fully masked.
For $QK^\top$, the mask determines which score
blocks are produced; for $PV$, the same mask determines which score
blocks participate in the reduction.  This optimization reduces the amount
of matrix work entering the pipeline without changing the tile-level
dependency structure.

\subsection{Layout-aware Runtime}
\label{sec:layout}

\noindent\textbf{Problem: layout compatibility gap.}
SME kernels prefer packed, tile-friendly layouts, but LLM runtimes often
store tensors in standard row-major or framework-native layouts.  A naive
runtime therefore packs inputs immediately before each SME kernel.  This
conversion adds memory traffic around the useful matrix computation and
places layout conversion on the critical path.  For decode FFN and
attention, layout conversion can take a similar amount of time as the SME
kernel itself, substantially reducing the net benefit of SME execution.

\noindent\textbf{Technique: layout-aware graph conversion.}
To reduce this overhead, \oursys{} tracks tensor layout as part of the
runtime state.  Each task declares which layouts it can consume and which
layout it produces.  The runtime then inserts conversions only when a
consumer requires a layout that the producer does not already provide.
\oursys{} uses two strategies to keep repeated packing off SME critical
paths.

\noindent\textbf{Off-critical-path packing for model weights.}
Model weights are static during inference, so their layout can be decided
before any request executes.  If the runtime determines that a weight
tensor will be consumed by SME kernels, \oursys{} packs it once at model
load time and stores the packed copy for reuse.  This avoids repacking the
same weights for every request and removes weight conversion from the
request critical path.

\begin{figure}[t]
\centering
\includegraphics[width=0.98\columnwidth]{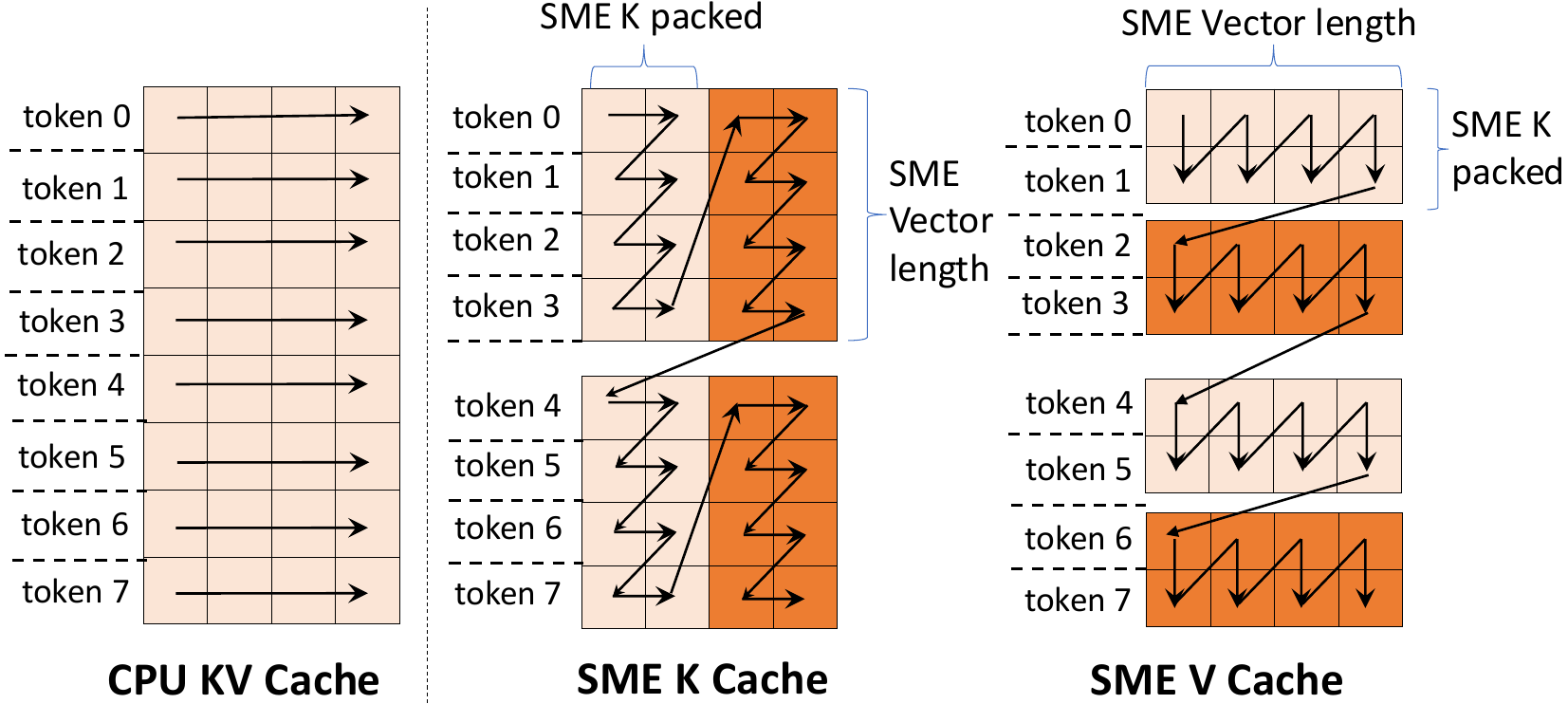}
\caption{Producer-side packing for SME-friendly KV cache layout.  A
standard CPU KV cache stores each token contiguously in row-major order.
\oursys{} stores K and V entries in SME-specific packed layouts at
production time, grouping elements according to the SME vector length and
the K-packed tile shape required by downstream attention kernels.}
\Description{The figure compares a standard CPU KV cache with two SME
layouts. The CPU KV cache stores token rows contiguously. The SME K cache
groups token blocks and K-packed columns for SME vector-length access. The
SME V cache stores token blocks in a different packed order for the V-side
attention access pattern.}
\label{fig:design-layout-conversion}
\end{figure}

\noindent\textbf{Producer-side packing for activation tensors.}
Activation tensors are generated online and
cannot be packed at model load time.  For these tensors, \oursys{} moves
conversion to the producer side when the downstream consumer is SME.
This removes an extra memory copy from the consumer path and,
for reused activations such as KV-cache entries, avoids repeatedly
unpacking the same data across attention invocations.
For example,
the key and value tensors produced by the
K-projection and V-projection are written directly into a packed
KV-cache layout. As shown in Figure~\ref{fig:design-layout-conversion},
a standard CPU KV cache stores each token as a contiguous row, while SME
attention prefers layouts that group tokens and feature elements according
to the SME vector length and kernel tile shape.  The runtime therefore stores
K and V cache entries in separate SME-friendly layouts.

Subsequent attention calls can then consume the
packed KV cache without repacking all previous entries on every invocation.

\subsection{Execution Plan Generation}
\label{sec:planner}

\begin{figure}[t]
\centering
\includegraphics[width=0.47\textwidth]{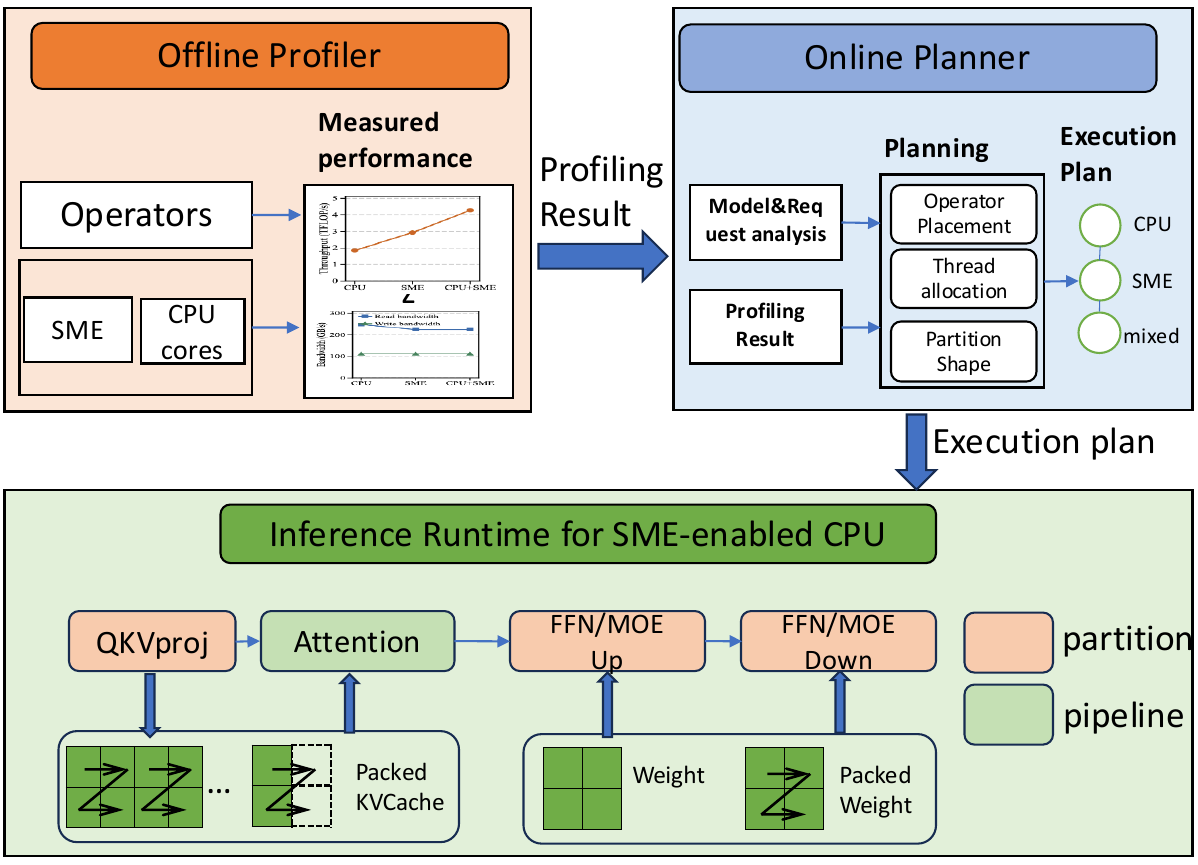}
\caption{The overall architecture of \oursys{}. The offline
profiler runs representative operators on SME and CPU cores to collect
hardware-specific performance metrics.  The online planner combines the
profiling result with model and request analysis to choose operator
placement, thread allocation, and partition shape.  The resulting plan
drives the inference runtime, including
the partition, pipeline, and packed-layout
mechanisms used by different LLM operators.}
\Description{The diagram shows an offline profiler, an online planner,
and an inference runtime. The offline profiler measures operator
performance on SME and CPU cores and passes the profiling result to the
online planner. The online planner uses model and request analysis plus
profiling results to decide operator placement, thread allocation, and
partition shape, producing CPU, SME, or mixed execution plans. The
runtime applies the plan to QKV projection, attention, FFN or MoE up and
down operators, packed KV caches, packed weights, partitioning, and
pipeline execution.}
\label{fig:design-architecture}
\end{figure}

Putting everything together, Figure~\ref{fig:design-architecture}
shows that \oursys{} generates an execution plan in two stages.  First,
an offline profiler measures the hardware capability of the target
SME-enabled CPU.  Second, an online planner combines these measurements
with the current request shape and the given model to produce a
tile-level execution plan.  The planner is invoked at operator
granularity, but the generated plan is executed at tile granularity so
that one operator can use SME, CPU cores, or both.  At runtime, the plan
selects the partitioned, pipelined, and packed-layout mechanisms used by
QKV projection, attention, and FFN/MoE operators.

\textbf{Performance Profiler}
The profiler executes a set of operator kernels on CPU cores and SME
to measure the performance metrics needed by the roofline-model analysis in Section~\ref{sec:sme}.
The performance metrics include:
(1) SME metrics: the SME unit count, matrix throughput, vector throughput
(2) CPU metrics: the CPU core count, per-core throughput for matrix and vector work.
(3) Memory metrics: the memory bandwidth and the cache size.
These metrics are only related to the hardware and are independent of the model and the request,
so they can be measured offline once and reused for different requests and models.

\textbf{Online Planner}
The online planner determines the optimal execution plan for each operator in
the computation graph, given the current request and the model.
It contains two processes.  The first is \emph{operator
placement}: for each operator, the planner decides whether it should run
on SME only, CPU cores only, or a mixed SME+CPU configuration.  The second
is \emph{tile-plan construction}: the planner decides the thread counts
and the partition size.

\noindent\textbf{Roofline-based operator placement.}
For an operator $o$, the planner first obtains its work $W_o$, useful
memory traffic $Q_o$, and shape parameters from the
operator graph and request state.  It then evaluates the roofline lower
bound in Section~\ref{sec:sme} for three execution targets: SME-only,
CPU-only, and mixed SME+CPU execution.  The SME-only target is preferred
when the operator has enough arithmetic intensity to amortize SME setup
and layout conversion, but does not expose enough independent output
tiles to keep CPU workers useful.  The CPU-only target is preferred for
memory-bound or vector-heavy operators, such as narrow decode GEMV and
softmax, where SME's matrix throughput cannot reduce the critical path.
The mixed target is preferred when the operator is compute-bound and has
enough independent output tiles for both SME and CPU cores.  This decision
is shape-dependent: the same logical operator may be CPU-only during
single-token decode, SME-only for a moderate-width decode shape, and mixed
SME+CPU during long-context prefill.

\noindent\textbf{Thread allocation.}
After operator placement, the planner chooses the number of SME and CPU
threads for the selected target.
\oursys{} uses a heuristic rather than an exhaustive search
because the choice is hardware-dependent, and the same request must be
planned quickly.  The heuristic first tries to cover the hardware units
selected by the placement decision.  It then checks whether the operator
exposes enough kernel-aligned tile regions for the candidate threads.  A
candidate thread is dropped if its assigned region would be smaller than
the native kernel tile or would force a non-aligned boundary.  This rule
avoids a common failure mode of over-partitioning: adding threads
increases scheduling overhead and may break the tile shapes that the SME
and CPU kernels were optimized for.

\noindent\textbf{Partition size.}
After choosing worker counts, the planner
computes the amount of output work assigned to SME and CPU cores.  Let
$P_{\mathrm{SME}}^{\mathrm{eff}}$ and
$P_{\mathrm{CPU}}^{\mathrm{eff}}$ be the effective throughputs measured by
the profiler for the selected kernel family and worker counts.  The ideal
fraction of the output dimension assigned to SME is
\begin{equation}
    \rho =
    \frac{P_{\mathrm{SME}}^{\mathrm{eff}}}
         {P_{\mathrm{SME}}^{\mathrm{eff}} +
          P_{\mathrm{CPU}}^{\mathrm{eff}}}.
    \label{eq:mixed-partition-ratio}
\end{equation}
The planner first computes the ideal split point $\rho D$, where $D$ is
the selected output extent.  It then rounds this split point to the
nearest boundary aligned with both the SME and CPU kernel tile sizes.  If
several aligned candidates are close to the ideal ratio, the planner
chooses the one that minimizes the predicted maximum of the SME-side and
CPU-side completion times.  As a result, the final split is not a fixed
FLOP ratio; it is the closest tile-aligned shape ratio supported by the
current kernel and thread allocation.

\section{Evaluation}
\label{sec:eval}

\subsection{Experimental Setup}

\begin{table*}[t]
\centering
\footnotesize
\caption{CPU platforms used in the evaluation.
}
\label{tab:eval-cpu-info}
\begin{tabular}{@{}p{0.22\textwidth}p{0.39\textwidth}p{0.15\textwidth}p{0.12\textwidth}@{}}
\toprule
Platform & CPU cores & SME:CPU ratio & Supported instruction sets \\
\midrule
Apple M4 Pro CPU &
14 cores (10 performance + 4 efficiency); 10 workers used &
2:10 (1 SME unit per 5 performance cores) &
SME, SME2 \\
MediaTek Dimensity 9500 &
8 cores (1 C1-Ultra + 3 C1-Premium + 4 C1-Pro) &
1:8 &
SME, SME2 \\
KunPeng 920 72F8 Server CPU &
server-class multi-core CPU; 10 workers used &
1:1 &
SME \\
\bottomrule
\end{tabular}
\end{table*}

\begin{table}[t]
\centering
\footnotesize
\caption{End-to-end benchmark configurations.}
\label{tab:e2e-datasets}
\begin{tabular}{@{}llcc@{}}
\toprule
Task type & Dataset &
\makecell{Mean prefill\\tokens} &
\makecell{Mean decoding\\tokens} \\
\midrule
Long-context QA & Ruler-4k/8k & 3324 / 7338 & 4 \\
Math QA & GSM8K & 92 & 100 \\
Long-form generation & LongWriter-6K & 339 & 5424 \\
\bottomrule
\end{tabular}
\end{table}

We implement \oursys{}, an optimized LLM inference engine for CPUs with
scalable matrix extensions.  \oursys{} is written in C and C++ and
includes SME-aware runtime components and SME kernels.  The SME kernels are
developed using the optimized assembly microkernels from
KleidiAI~\cite{arm2026kleidiai}
and Arm C Language Extensions (ACLE) SME
intrinsics~\cite{arm2026acle}.
We evaluate the performance of \oursys{} on three platforms equipped with
CPUs supporting Scalable Matrix Extension (SME): Apple M4 Pro CPU,
MediaTek Dimensity 9500, and KunPeng 920 72F8 Server CPU with SME support.
These are recently released CPU chips that support SME or SME2, and
they represent the current state of the art for SME performance.
Table~\ref{tab:eval-cpu-info} summarizes their CPU resources, SME-to-CPU
resource ratio, and supported instruction sets.

In our evaluation,
we compare \oursys{} against llama.cpp~\cite{gerganov2023llamacpp}, one of the most widely used CPU LLM inference framework.
We use the default CPU backend for all platforms.
We select both dense and MoE models for evaluation:
Llama-3.2-3B~\cite{meta2024llama32},
Qwen3-4B~\cite{yang2025qwen3technical}, and
Qwen3-30B-A3B~\cite{yang2025qwen3technical}.
The Qwen3-30B-A3B model uses a 4-bit weight quantization
to fit into device memory.
We evaluate
three representative tasks: long-text processing~\cite{hsieh2024ruler}, simple question answering~\cite{cobbe2021gsm8k},
and long-text generation~\cite{bai2024longwriter}, as shown in Table~\ref{tab:e2e-datasets}.  The Ruler row
sweeps sequence lengths of 4K and 8K, while GSM8K and LongWriter-6K sweep
batch size.
We use 10 threads and 8 threads for Apple M4 Pro and MediaTek Dimensity 9500,
respectively,
which equals the number of performance cores on the two platforms.
For the KunPeng 920 Server CPU, we use 10 threads for a fair comparison.

\subsection{End-to-End Inference}

\begin{figure*}[t]
    \centering
    \includegraphics[width=0.98\textwidth]{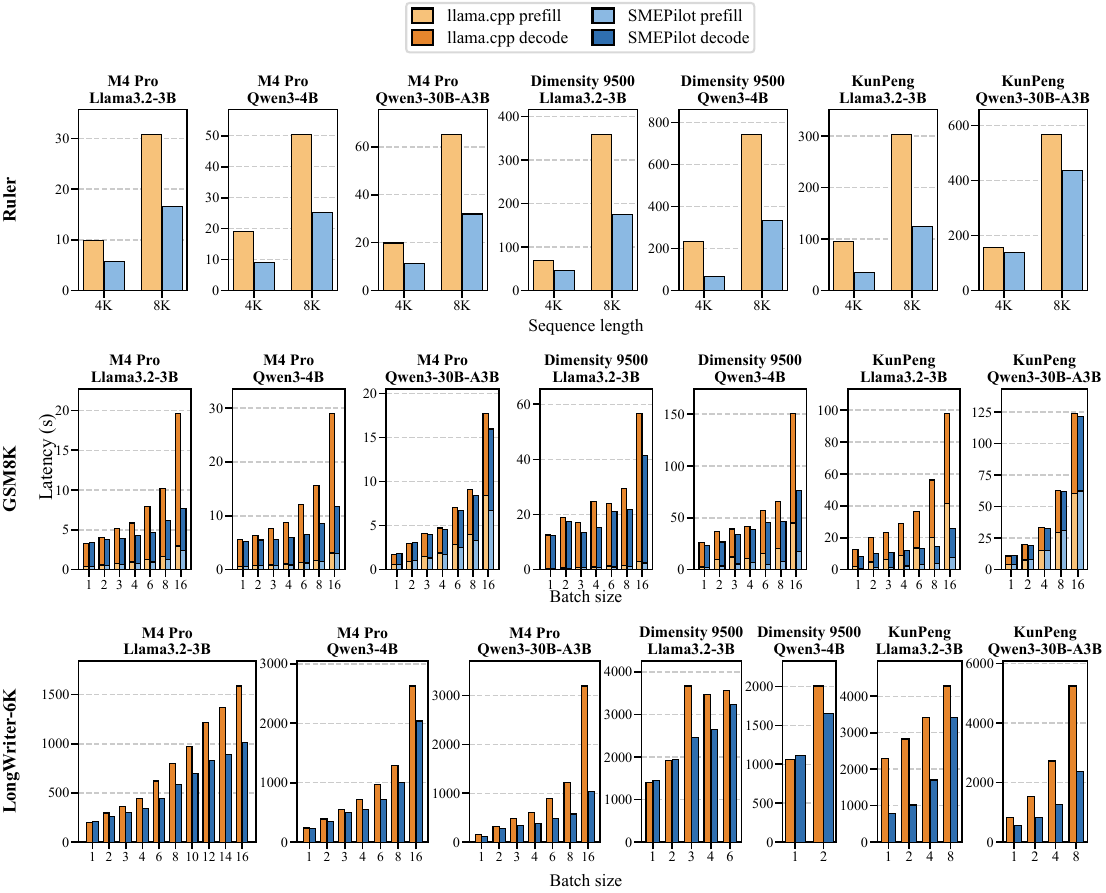}
    \caption{End-to-end latency across devices and real-world datasets.
    Rows correspond to Ruler, GSM8K, and LongWriter-6K. Within each row,
    panels compare the measured device-model pairs: Apple M4 Pro with
    Llama3.2-3B, Qwen3-4B, and Qwen3-30B-A3B; MediaTek Dimensity 9500
    with Llama3.2-3B and Qwen3-4B; and KunPeng Server CPU with
    Llama3.2-3B and Qwen3-30B-A3B. We skip some batch or model-pair
    because of the memory constraints of device.}
    \Description{A multi-panel bar chart comparing baseline and SMEPilot
    end-to-end latency. The rows are Ruler, GSM8K, and LongWriter-6K.
    Each row contains measured device-model panels for Apple M4 Pro,
    MediaTek Dimensity 9500, and KunPeng Server CPU. Ruler uses sequence
    length on the x-axis, and the other rows use batch size.}
    \label{fig:end-to-end}
\end{figure*}

Figure~\ref{fig:end-to-end} shows the end-to-end inference latency of
\oursys{} compared to the llama.cpp baseline on the CPU platforms.
\oursys{} produces up to 3.94 $\times$ speedup across different datasets and CPU platforms.
More specifically, in the prefill stage, \oursys{} achieves 1.13$\times$ to 5.42$\times$ speedup.
In the decode stage, \oursys{} produces a up to 3.48$\times$ speedup.

In the long text processing task,
\oursys{} produces a 1.13$\times$ to 3.56$\times$ speedup on three different CPUs.
The speedup arises from both utilizing CPU cores and SME units together for compute-bound prefill operators,
including the attention, FFN and MoE.

In the simple question answering task,
\oursys{} produces a 1.0$\times$ to 3.94$\times$ speedup on three different CPUs.
The speedup arises from our roofline-based operator placement for
prefill, decode phases and different batch sizes.
For batch sizes of 1 or 2, \oursys{} produces a comparable performance with the baseline
because the operators are more memory-bound and the runtime selects CPU execution.
When the batch size increases, \oursys{} produces a more significant speedup.

This is because larger batch sizes reach
the ridge-crossing regime where SME can be more effective,
which demonstrates the effectiveness of our online planner.
The speedup is smaller for the MoE models. This is because
tokens are routed to different experts, which reduces the effective batch size for each expert and
makes the operators more memory-bound, thus reducing the benefit of using SME.

In the long-text generation task, \oursys{} produces a 1$\times$ to 2.95$\times$ speedup.
This long generation task has a much longer decoding phase,
and \oursys{} accelerates the decode phase by using SME for operators in the ridge-crossing regime.

Notably, the speedup is more significant for the MoE model in this task.
This is because Qwen3-30B-A3B uses group-query-attention with larger group size,
which increases the arithmetic intensity of attention and help
gain more speedup from our SME-aware scheduling.

\subsection{Operator Performance Breakdown}

To better analyze the end-to-end speedup and demonstrate
the effectiveness of our design,
we conduct a detailed tests for operators in the prefill and decode phases
on the Apple M4 Pro CPU.

\subsubsection{FFN}

\begin{figure}[t]
    \centering
    \includegraphics[width=\columnwidth]{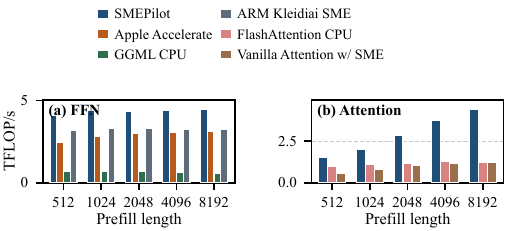}
    \caption{Prefill operator throughput on the Apple M4 Pro CPU.
    (a) FFN GEMM for the $B \times 3072 \times 8192$ shape, compared with
    Apple Accelerate, GGML CPU, and ARM Kleidiai SME baselines.
    (b) Attention for Llama-3.2-3B with 24 query heads, 8 KV heads, and
    head dimension 128, compared with FlashAttention CPU and vanilla
    attention with SME.}
    \label{fig:prefill-ops}
\end{figure}

Figure~\ref{fig:prefill-ops}(a) shows the prefill FFN GEMM performance
of \oursys{} compared with Apple Accelerate~\cite{apple2026accelerate},
GGML~\cite{gerganov2023ggml} CPU, and the ARM
Kleidiai SME~\cite{arm2026kleidiai}, across different prefill lengths.
For prefill length from 512 to 8192,
\oursys{} sustains 4.10--4.44~TFLOP/s.
Compared with the GGML CPU, \oursys{} achieves a 6.06--7.43$\times$ speedup
across different prefill lengths, demonstrating the effectiveness of our roofline analysis.
Compared with the Apple Accelerate and ARM Kleidiai SME,
which uses the SME unit but does not use \oursys{}'s mixed execution
plan, \oursys{} still achieves a 1.30--1.67$\times$ speedup.  The
performance improvement arises from using the SME unit and CPU cores
together (Section~\ref{sec:sme}) and from the partition technique that
balances the two sides (Section~\ref{sec:design}).

\subsubsection{Prefill Attention}

Figure~\ref{fig:prefill-ops}(b) shows the prefill attention performance
of \oursys{} compared to the CPU FlashAttention and vanilla
attention with SME across different sequence lengths.
For attention with 24 query heads and 8 KV heads, \oursys{}
sustains 1.50--4.38~TFLOP/s as the prefill length increases from 512 to
8192, and achieves a 1.56--3.50$\times$ speedup over the FlashAttention CPU baseline.
The gain demonstrates the effectiveness of our phase-aware pipelining design
in Section~\ref{sec:design}.

\subsubsection{Decode FFN}

\begin{figure}[t]
    \centering
    \includegraphics[width=\columnwidth]{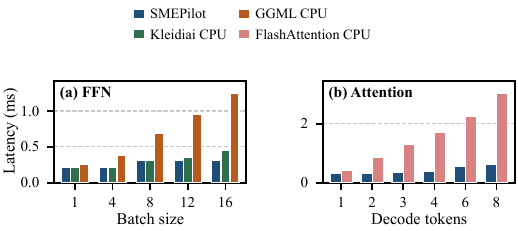}
    \caption{Decode operator latency on the Apple M4 Pro CPU.
    (a) FFN GEMV for the $2560 \times T \times 9728$ shape, compared
    against Kleidiai CPU and GGML CPU baselines. (b) Attention with KV
    length 8192, 32 query heads, 8 KV heads, and head dimension 128,
    compared with the FlashAttention CPU baseline.}
    \label{fig:decode-ops}
\end{figure}

Figure~\ref{fig:decode-ops}(a) reports the decode FFN result.
For different decode batch sizes,
\oursys{} achieves a comparable performance with the CPU baselines when the batch size is small,
and achieves 1.45$\times$ and 3.95$\times$ speedup
compared to the Kleidiai CPU and GGML CPU respectively when the batch size is large.
This is because the runtime selects CPU execution for smaller batches,
but selects SME for larger batches that reach the ridge-crossing regime.

\subsubsection{Decode Attention}

Figure~\ref{fig:decode-ops}(b) shows the latency of decode attention of \oursys{} and
the FlashAttention CPU baseline.
\oursys{} improve decode attention latency from 1.32$\times$ to 4.75$\times$ depending on the decode token width.
The benefit arises from the runtime's ability to use SME when the decode token width is large enough
and to pipeline matrix and vector phases when dependencies permit.
The speedup grows with the decode token width because wider decode attention has larger matrix shapes
and has higher arithmetic intensity,
making it more profitable to use SME and to pipeline matrix and vector phases.
This result validates the design choice of treating decode attention as a shape-dependent case
instead of a purely memory-bound operator that should always run on cpu cores.

\subsection{Ablation Study}

\begin{figure}[t]
    \centering
    \includegraphics[width=\columnwidth]{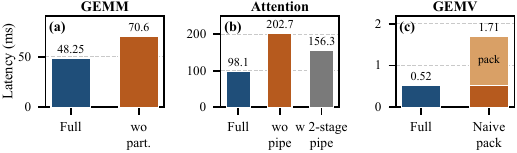}
    \caption{Ablation study on the Apple M4 Pro CPU.  Tile partition,
    phase-aware pipelining, and layout-aware execution reduce GEMM(4096x3072x8192),
    prefill attention(seq 4096), and decode GEMV(16x3072x8192) latency, respectively.}
    \Description{Three bar-chart panels report latency for GEMM,
    attention, and GEMV ablations.  Full SMEPilot is faster than the
    no-partition, no-pipeline, two-stage pipeline, and naive-packing
    variants.}
    \label{fig:eval-ablation}
\end{figure}

Our ablation study focuses on the three key design mechanisms in
Section~\ref{sec:design}: tile-level work partitioning, phase-aware
pipeline execution, and layout-aware execution.  As shown in
Figure~\ref{fig:eval-ablation}, removing tile partitioning increases GEMM
latency by 1.46$\times$ because the operator can no longer use SME and CPU cores together.
Removing the attention pipeline increases
prefill attention latency by 2.07$\times$.
Finally, naive on-path layout packing
increases decode GEMV latency from 0.52~ms to 1.71~ms, confirming that
layout conversion must be kept off repeated SME critical paths.

\subsection{Power Consumption}

\begin{table}[t]
\centering
\footnotesize
\caption{End-to-end energy on Apple M4 Pro for Qwen3-4B
running the Ruler-4K workload.}
\label{tab:energy-ruler}
\setlength{\tabcolsep}{2.4pt}
\begin{tabular}{@{}lrrrr@{}}
\toprule
Mode & $n$ & \makecell{Throughput\\(tok/s)} &
\makecell{avg\\W} &
\makecell{E2E energy (J)} \\
\midrule
llama.cpp baseline & 3 & $188.812 \pm 7.787$ & $26.050 \pm 2.193$ & $482.813 \pm 23.213$ \\
\oursys{} & 3 & $338.012 \pm 0.680$ & $22.611 \pm 0.296$ & $233.931 \pm 2.751$ \\
\bottomrule
\end{tabular}
\end{table}

We further run a power measurement on the Mac platform using Qwen3-4B
and the Ruler-4K workload.  We choose this platform because macOS provides
the \texttt{powermetrics} tool for sampling system power during inference.
Table~\ref{tab:energy-ruler} reports the end-to-end energy
from this measurement.  \oursys{} reduces the energy to
0.485$\times$ of the llama.cpp baseline.  The
lower energy mainly comes from finishing inference faster and better utilize SME.

\subsection{GPU Inference Comparison}

\begin{figure}[t]
\centering
\includegraphics[width=\columnwidth]{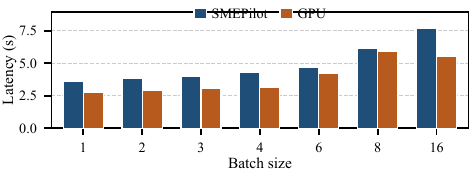}
\caption{End-to-end Llama3.2-3B GSM8K latency on Apple M4 Pro, comparing \oursys{} on
the CPU with GPU inference across batch sizes.}
\Description{A grouped bar chart comparing SMEPilot and GPU end-to-end
latency on GSM8K for batch sizes 1, 2, 3, 4, 6, 8, and 16.}
\label{fig:gpu-comparison}
\end{figure}

We also compare \oursys{} with GPU inference on the Apple M4 Pro platform
because Macbook Pro is equipped with a 20-core GPU with 8 TFlops.
As shown in Figure~\ref{fig:gpu-comparison}, \oursys{} achieves
0.72--0.96$\times$ performance compared to the GPU inference.
This result positions \oursys{} as a practical CPU
inference engine for systems where the GPU is unavailable or reserved for
other workloads, or to coordinate with GPU execution.

\section{Related Work}
\label{sec:rel}

\subsection{CPU LLM Inference}

CPU inference has become an important deployment path for LLMs because
CPUs are widely available across edge devices, personal computers, and
servers.  Practical runtimes such as llama.cpp~\cite{gerganov2023llamacpp}
show that careful low-level kernels, weight quantization, and CPU-friendly
memory layouts can make local LLM inference usable on commodity machines.
Intel Extension for Transformers studies CPU LLM inference on Xeon
processors, combining weight-only quantization, operator fusion, memory
management, and runtime optimizations for popular LLMs~\cite{shen2023efficientcpu}.
ArcLight further studies lightweight LLM inference on many-core CPUs~\cite{xu2026arclight}.
Other CPU-oriented systems target low-bit or sparse execution: T-MAC
replaces mixed-precision GEMM with table lookup for edge CPUs~\cite{wei2025tmac},
while sparse Transformer inference systems reduce CPU-side compute and
memory traffic~\cite{ps2024cpuinference}.  These systems demonstrate that
CPU LLM inference is practical, but their main target is model compression,
quantized arithmetic, or CPU-only kernel optimization.

Several works further analyze CPU inference as a phase-dependent systems
problem.

Sandwich observes that prefill and decode have different
execution characteristics and searches different CPU configurations for the
two phases~\cite{zhao2025sandwich}.  Platform studies show that LLM
inference stresses memory bandwidth, scheduling, and distributed inference
resources differently across model sizes and request regimes~\cite{bambhaniya2024demystifying}.
On Arm CPUs, prior work characterizes Transformer inference on many-core
processors~\cite{jiang2022characterizing} and optimizes attention by
exploiting data reuse across Arm multi-core CPUs~\cite{fu2024optimizing}.
Other systems use CPUs as partners in heterogeneous LLM inference.  FlexGen
offloads tensors across GPU, CPU, and storage~\cite{sheng2023flexgen};
PowerInfer splits work between GPU and CPU on commodity PCs~\cite{song2024powerinfer};
PowerInfer-2 coordinates NPU, CPU, and storage on smartphones~\cite{xue2024powerinfer2};
and Toppings uses CPU computation to hide LoRA adapter loading during LLM
serving~\cite{li2025toppings}.

\oursys{} is complementary to these CPU
inference and LLM serving systems~\cite{romero2021infaas}.  Rather than scheduling requests across
devices, or optimizing a conventional
vector-core runtime, \oursys{} targets CPUs that expose both
matrix-extension units and ordinary CPU cores.  It therefore focuses on
when to use SME, when to keep work on CPU cores, and when to coordinate both
within one operator.

\subsection{CPU Matrix Extension and SME}

Modern CPUs increasingly include matrix extensions that provide
accelerator-like matrix throughput inside the CPU package.  Intel AMX
adds tile registers and tile matrix instructions, and prior work has used
AMX to accelerate LLM inference on CPUs~\cite{kim2024amxllm}.  SparAMX
combines AMX with unstructured sparsity to accelerate compressed LLM token
generation on Intel CPUs~\cite{abouelhamayed2025sparamx}.  LIA uses
AMX-enabled CPU computation as part of a CPU-GPU-CXL cooperative inference
system~\cite{kim2025lia}.  These works show that CPU matrix extensions can
be useful for LLM workloads, especially when the software stack exposes
enough matrix work to the extension.  However, they primarily study Intel
AMX, compressed models, or CPU-GPU offload.  \oursys{} instead studies Arm
SME and the scheduling problem inside an SME-enabled CPU cluster, where
SME and CPU cores share cache and memory bandwidth.

Arm SME introduces matrix-oriented architectural state and outer-product
instructions on top of the Arm vector execution model~\cite{arm2024smeintro}.
Recent SME studies characterize the extension and build optimized kernels
for dense matrix multiplication~\cite{remke2024hellosme,deng2025demystifying}
and sparse matrix multiplication~\cite{lei2025loops}.  These kernel-level
studies are an important foundation for understanding SME's peak
throughput and data layout requirements.  \oursys{} builds on this line
of work but asks a different systems question: how should an LLM runtime
use SME across prefill, decode, attention, FFN, vector operations, and
layout-changing operator boundaries?  The answer is not to replace every
kernel with an SME kernel.  \oursys{} uses a roofline-guided planner to
select SME-only, CPU-only, or mixed SME+CPU execution; partitions
compute-bound operators across heterogeneous units; pipelines
SME-friendly matrix phases with CPU-friendly vector phases; and tracks
layouts to keep packing out of repeated SME critical paths.

\section{Discussion and Future Work}
\label{sec:discussion}

\noindent\textbf{Design implications for future CPU matrix extensions.}
Based on our characterization in using SME for LLM inference,
we suggest that current SME can be improved along following dimensions.
First, stronger support for vector-heavy and
GEMV-like phases is important.  Decode, softmax, and
normalization do not naturally exploit SME outer-product throughput, and
direct SME execution can be slower than using CPU cores.  Increasing
matrix throughput alone therefore will not fully accelerate LLM inference
unless the architecture or runtime also improves these non-matrix phases
or makes them easier to overlap with matrix work.

Second, matrix extensions need more convenient layout management.  SME
kernels require packed, tile-friendly layouts, while the surrounding LLM
pipeline often uses conventional tensor layouts.  Without architectural or
runtime support for cheap conversion, layout traffic can erase much of the
gain from a faster matrix kernel.

Third, future systems would benefit from clearer isolation and
between SME units and CPU cores.  Currently, SME and CPU
cores can provide near-additive compute throughput, but they still contend
for cache capacity, and the CPU cores that issue SME
instructions.  A more decoupled SME from cpu pipeline
would reduce contention.

\noindent\textbf{Cooperation with GPU and NPU.}
While this paper focuses on CPU with scalable matrix extensions, future LLM inference
systems will likely need to coordinate across CPU, GPU, and NPU execution.
CPU matrix extensions are attractive because they are integrated into
general-purpose CPUs, which are widely deployed and cost-effective across
both edge and server platforms, and can execute CPU-side work without an
additional device transfer.  GPUs and NPUs, in contrast, provide much
higher throughput for large regular matrix workloads.  A future inference
runtime could therefore use GPUs or NPUs together with CPU matrix extensions,
exposing more overlap opportunities.  Realizing
this cooperation requires a broader planner that accounts for factors,
such as device transfer cost, memory placement and etc.  We leave such
cooperation with SME to future work.

\section{Conclusion}
\label{sec:conclusion}

This paper studies how Arm SME should be used for LLM inference.  Our
characterization shows that SME is not a universal substitute for CPU vector
cores.  \oursys{} uses this
observation to select SME-only, CPU-only, or mixed execution, partition work
at tile granularity, overlap matrix and vector phases, and avoid repeated
layout conversion.  The resulting speedups show that CPU matrix extensions
are most effective when their throughput is applied selectively and
amortized across suitable inference regimes.

\clearpage

\bibliography{papers}

\end{CJK*}
\end{document}